\documentclass[letterpaper, 10 pt, conference]{ieeeconf}  

\IEEEoverridecommandlockouts                              

\usepackage{graphics} 
\usepackage{epsfig} 
\usepackage{mathptmx} 
\usepackage{times} 
\usepackage{amsmath} 
\usepackage{amssymb}  

\usepackage[english]{babel}
\usepackage[T1]{fontenc}
\usepackage[utf8]{inputenc}

\usepackage{amsfonts,amsmath,bm,hhline}
\usepackage{epsfig}
\usepackage{color}
\usepackage[hang]{caption2}
\usepackage[normalsize,it,hang]{subfigure}
\newtheorem{remark}{Remark}  
  
\DeclareGraphicsExtensions{.jpg,.png,.jpg,.jpg}

\usepackage{algorithm}
\usepackage{algpseudocode}

\title{{\tt \scriptsize European Control Conference (ECC) --- July 7-10, 2026, Reykjavík, Iceland} \\ \LARGE {\bf A control-theoretic simplification of \\ adaptive bitrate (ABR) video streaming}}

\author{ Michel Fliess$^{1,3}$ and C\'edric Join$^{2}$
\thanks{$^{1}$ LJLL (CNRS, UMR 7598), Sorbonne Universit\'{e}, 4 place Jussieu, 75005 Paris, France (e-mail: michel.fliess@sorbonne-universite.fr, michel.fliess@swissknife.tech).}%
\thanks{$^{2}$CRAN (CNRS, UMR 7039)), Universit\'{e} de Lorraine, BP 239, 54506 Vand{\oe}uvre-l\`{e}s-Nancy, France (e-mail: cedric.join@univ-lorraine.fr)}%
\thanks{$^{3}$ AL.I.E.N., 7 rue Maurice Barr\`{e}s, 54330 V\'{e}zelise, France (e-mail: michel.fliess@alien-sas.com)}%
}

\begin{document}

\maketitle
\thispagestyle{empty}
\pagestyle{empty}

\begin{abstract}
Adaptive bitrate streaming (ABR) over the HyperText Transfer Protocol (HTTP), which raises numerous delicate questions,
is nowadays almost the only approach to video streaming. This paper presents elementary solutions to three key issues:
1) A straightforward feedforward control strategy for the bitrate and the buffer level via flatness-based control.
2) Closing the loop permits mitigating unavoidable mismatches and disturbances, such as Internet fluctuations. This is adapted from the new HEOL setting, which mixes model-free and flatness-based controls.
3) An easily implementable closed-form estimate of the bandwidth via algebraic identification techniques is
derived, perhaps for the first time. It permits handling severe variations
in channel capacity. Several computer experiments and metrics for evaluating the Quality of Experience (QoE) are displayed and discussed.    
\end{abstract}


\section{Introduction}
As predicted more than twenty years ago \cite{hellerstein}, advanced control-theoretic tools are playing a r\^{o}le in computing systems, as recently demonstrated for avoiding Internet congestion \cite{mounier}. This paper is devoted to video streaming, where \emph{adaptive bitrate} (\emph{ABR}) {\em streaming}, which raises numerous delicate questions, is almost the only approach employed in practice (see, {\it e.g.}, \cite{hard,qin0,qin,qin.phd}). The following facts are  highlighted in \cite{qin0,qin,qin.phd}:
\begin{enumerate}
    \item The criticisms (see, \textit{e.g.}, \cite{tian,yin}) against PIDs in ABR do not withstand closer scrutiny.
    \item The recent buffer-based approach \cite{bba} turns out to employ a simplified PID.
    \item The PID controllers developed in \cite{qin0,qin,qin.phd} yield quite convincing results when compared to the ABR algorithms in \cite{yin}, \cite{bba} (model-predictive control) and \cite{bola} (Lyapunov analysis and optimal control).
\end{enumerate}
\begin{remark}
See, e.g., \cite{feed,cofano,transc,hoss,amazon} for some other control settings.
\end{remark}
In spite of their huge industrial popularity, PIDs not only exhibit severe difficulties with gain tuning but also a lack of robustness (see, \textit{e.g.}, \cite{dwyer}). This is why we suggest replacing them with new tools:
\begin{enumerate}
    \item There is a well-accepted model via nonlinear differential equations (see, \textit{e.g.}, \cite{qin0,qin,qin.phd}), where the control and output variables are, respectively, the bitrate $R(t)$ and the buffer level $x(t)$. This system is (\emph{differentially}) \emph{flat} and $x(t)$ is a \emph{flat output} \cite{flmr_ijc,flmr_ieee}. This property, which makes it easy to assign an open-loop reference trajectory, {\it i.e.}, a feedforward control strategy, has already been exploited many times: see the books \cite{sira1,levine,rudolph} for concrete engineering examples.
    \item The feedback law is adapted from the new \emph{HEOL}\footnote{HEOL: ``Sun'' in the Breton language.} setting \cite{heol}, which recasts \emph{model-free control} \cite{mfc1,mfc2} and sheds new light on flatness-based control. Robustness with respect to severe model mismatches and external disturbances is ensured (see \cite{wien,degorre,mfpc,maroc}). It fits well with piecewise constant control variables, which can only take a finite number of values \cite{toulon,dirn}.
\end{enumerate} 
The channel or network capacity is a time-varying parameter of utmost importance. 
In spite of many references regarding its estimation (see, \textit{e.g.}, \cite{prasad,zou}), to the best of our knowledge, there is no connection to the vast control literature on estimation and identification. This parameter turns out to be ({\em rationally}) {\em identifiable} \cite{easy,sira2}: it is a rational function of $R(t)$ and $x(t)$ and their derivatives up to some finite order:
\begin{itemize}
   \item It results, perhaps for the first time, in an easily implementable closed-form expression for a bandwidth estimate (see, \textit{e.g.}, \cite{wang} in a completely different area).    
   \item It should be most helpful to take into account wild Internet fluctuations (see, \textit{e.g.}, \cite{huang}). Compare with \cite{qin0,qin,qin.phd} where $\mathfrak{u}(t) = \frac{R(t)}{C(t)}$ is the control variable: this makes it difficult to account for bandwidth variations.
    \item We are able to track trajectories for the buffer level, which might improve the \emph{Quality of Experience} (\emph{QoE}) (see, \textit{e.g.}, \cite{qoe}).
\end{itemize}
\begin{remark}
    Compare with attempts via {\em deep reinforcement learning} ({\em DRL}) (see, e.g., \cite{drl,drl1}, and references therein).
\end{remark}

Our paper is organized as follows. Sect. \ref{model} shows that the modeling in \cite{qin0,qin,qin.phd} is flat. The bandwidth rational identifiability is exploited in Sect. \ref{bandwidth} to derive a closed-form expression for its estimation. Computer experiments and metrics for evaluating the QoE are presented in Sect. \ref{simu}. Sect. \ref{conclu} contains some hints for future investigations.


\section{Conceptual tools}\label{model}
\subsection{Flatness}
\subsubsection{Definition}
Consider, for simplicity's sake, a control system with a single control variable. It is said to be (\emph{differentially}) \emph{flat} \cite{flmr_ijc,flmr_ieee} if, and only if, there exists a variable $z$ satisfying the next two properties: 1) $z$ can be expressed as a {\em differential function} of the system variables, \textit{i.e.}, a function of those variables and their derivatives up to some finite order; 2) any system variable can be expressed as a differential function of $z$. The variable $z$ is called a \emph{flat output}.
\subsubsection{Equations}
Introduce \cite{qin0,qin,qin.phd}
\begin{equation}
\label{1}
    \dot{x} (t) = \frac{C (t)}{R (t)} \quad {\rm if} \ t < \delta \ {\rm and/or} \ x(t) < \Delta
\end{equation}
and
\begin{equation}
\label{2}
    \dot{x} (t) = \frac{C (t)}{R (t)} - 1 \quad {\rm if} \ t \geq \delta \ {\rm and} \ x(t) \geq \Delta
\end{equation}
The control variable $R(t)$ is the bitrate of the \emph{chunk}, \textit{i.e.}, a video segment that is downloaded at time $t$. The output variable $x(t)$ is the client buffer level (in seconds). The network bandwidth is $C(t)$, and $\Delta$ (resp. $\delta$) is the chunk duration (resp. startup delay). See \cite{qin0,qin,qin.phd} for more details, including block diagrams. Eq. \eqref{1} and \eqref{2} define flat systems: $R(t)$ is a flat output, i.e.,
$R(t) = \frac{C(t)}{\dot{x}(t)}$ \text{for Eq. \eqref{1}}, 
and
$R(t) = \frac{C(t)}{\dot{x}(t) + 1}$ for Eq. \eqref{2}.
\begin{remark}
If $t > \delta$ and $x(t) < \Delta$, there is a video \emph{stall} or \emph{rebuffering}: the video is ``frozen.'' This key degradation is taken into account by the third metric in Sect. \ref{metric}.
\end{remark}

\subsection{HEOL}
\subsubsection{General principles}
Differentiating Eqs. \eqref{1} and \eqref{2} yields $d\dot x =\frac{dC(t)R(t)-C(t) dR(t)}{R^2(t)}$.
Introduce, according to \cite{heol}, the \emph{homeostat}
\begin{equation}\label{hom}
    \frac{d}{dt} (\Delta x) = \mathfrak{F} + \alpha \Delta R
\end{equation}
where 
\begin{itemize}
    \item $\Delta R = R - R^\star$, $\Delta x = x - x^\star$ ($R^\star$, $x^\star$ are the flatness-based open-loop references);
    \item $\alpha (t) = - \frac{C^\star(0)}{(R^\star(t))^2}$.\footnote{ Notice that we did not write $\alpha (t) = - \frac{C^\star(t)}{(R^\star (t))^2}$ since $C^\star(t)$ is the network bandwidth, which should be continuously estimated (see Sect. \ref{bandwidth}). This important discrepancy with \cite{heol} is thus due to technological constraints.} which are generally time-varying; 
    \item $\mathfrak{F}$ subsumes the poorly known system structure and the disturbances and therefore is not necessarily equal to
    $\frac{\dot{C}(t)}{R(t)}$. 
\end{itemize}
Mimicking \cite{mfc1}, associate \cite{heol} with the homeostat \eqref{hom} the \emph{intelligent proportional} (\emph{iP}) controller
\begin{equation}\label{ip}
 \Delta R = - \frac{\mathfrak{F}_{\rm est} + K_{P} \Delta x}{\alpha}   
\end{equation}
where $K_{P} \in \mathbb{R}$ is the gain and $\mathfrak{F}_{\rm est}$ is an estimate of $\mathfrak{F}$ (see Sect. \ref{estimat}). Combine Eqs. \eqref{hom} and \eqref{ip}:
 $\left(\frac{d}{dt} + K_{P}\right) \Delta x  = \mathfrak{F} - \mathfrak{F}_{\rm est}$. Thus, if the estimate is ``good,'' i.e., $\mathfrak{F} - \mathfrak{F}_{\rm est} \approx 0$,
 $\lim_{t \to \infty} \Delta x \approx 0$ if and only if $K_{P} >0$.

\subsubsection{Discontinuous aspects}
The fact that the bitrate is piecewise constant and takes only a finite number of values implies that the iP \eqref{ip} should be replaced by
\begin{equation}\label{ip.disc}
\Delta R = - \frac{\mathfrak{F}_{\rm est} + K_{P} \Delta x}{\alpha}  + \epsilon
\end{equation}
where $\epsilon (t)$ is bounded, \textit{i.e.}, $|\epsilon (t)|  < B$, $B > 0$. Then
\begin{equation*}\label{dyna2}
\dot{e} + K_P e = \mathfrak{F} - \mathfrak{F}_{\text{est}} + \alpha \epsilon
\end{equation*}
If the estimate $\mathfrak{F}_{\text{est}}$ is good again, it is well known that the tracking error $e(t)$ remains bounded if and only if $K_P > 0$ (classic Bounded Input Bounded Output (BIBO) stability).


\subsection{Bandwidth estimation}\label{bandwidth}
\subsubsection{Identifiability}
Eq. \eqref{1} (resp. \eqref{2}) yields $C(t) = R(t) \dot{x}(t)$ (resp. $C(t) = R(t) (1 + \dot{x}(t))$). The parameter $C(t)$ is a rational function of the control variable $R(t)$ and the derivative of the output variable $x(t)$. It is said to be {\em rationally identifiable} \cite{easy,sira2}.
\subsubsection{Estimation}\label{estimat}
Examine only Eq. \eqref{2} for the sake of simplicity. Remember that $R(t)$ is piecewise constant. It is also possible to approximate $C(t)$ by a piecewise constant function, i.e., by a function that is constant over small time intervals (see, e.g., \cite{bourbaki}). Rewrite Eq. \eqref{2} accordingly:
\begin{equation*}\label{cst}
\mathcal{C} = \mathcal{R} (\dot{\mathrm{x}}(t) + 1) 
\end{equation*}
where $\mathcal{C}$, $\mathcal{R}$ are real constants, and $\mathrm{x}(t)$ is a time function. Bandwidth estimation boils down to the estimation of $\mathcal{C}$ from the right hand side of the above equation. Classic operational calculus (see, \textit{e.g.}, \cite{yosida}) yields
\begin{equation}\label{operat}
\frac{\mathcal{C}}{s} = \mathcal{R} \left(s\mathrm{X} - \mathrm{x}(0) + \frac{1}{s}\right)
\end{equation}
where $s$ (resp. $\mathrm{X}$) is often called the Laplace variable (resp. the Laplace transform of $\mathrm{x}$). Mimic calculations in \cite{mfc1}. 
In order to get rid of $\mathrm{x}(0)$ take the derivatives with respect to $s$ of both sides of Equation \eqref{operat}. Then multiply both sides by $s^{-2}$ in order to smooth out corrupting noises: 
$
\frac{\mathcal{C}}{s^4} = \mathcal{R}\left(\frac{X}{s^2} + \frac{1}{s} \frac{\frac{dX}{ds}}{s} - \frac{1}{s^4} \right)
$.
The equivalence of $\frac{d}{ds}$ with the product by $-t$ in the time domain yields 
\begin{equation*}\label{interm}
\mathcal{C} =-\frac{6 \mathcal{R}}{t^3} \left(\int_0^t(t-2\sigma)\mathrm{x}(\sigma) d\sigma +1\right)
\end{equation*}
Assuming that $C(t)$ and $R(t)$ are approximately constant on the interval $[t-\tau, t]$ yields the following bandwidth estimate
\begin{equation}\label{estim}
C_{\rm est} (t) =-\frac{6 R(t)}{\tau^3} \left( \int_{0}^\tau (\tau-2\sigma )x(t - \tau + \sigma ) d\sigma 
+1\right)
\end{equation}
where $\tau > 0$, is ``small.'' Formula \eqref{estim}, which makes sense if $t > \delta + \tau$ and $x(t) > \Delta$, is a low pass filter. 

\section{Simulation issues}\label{simu}
\subsection{Preliminary considerations: Flatness-based open-loop}
The numerical values of $R(t)$ belong to a finite set $\mathfrak{U}$. $C(t) = C(0)$ is assumed, for the sake of simplicity, to be constant. A reference trajectory $x^\star (t)$ is deduced from a classic B\'{e}zier curve (see, e.g., \cite{rogers}):
\begin{equation*}
    \begin{cases}
        x^\star(t)=&x^\star(t_0)\text{ if } 0\leq t \leq t_0\\
        x^\star(t)=&x^\star(t_0)+(x^\star(t_f)-x^\star(t_0))T^4\\
        &(70-224T+280T^2-160T^3+35T^4) \\ 
        & \text{if } t_0 \leq t \leq t_f \\
        x^\star(t)=&x^\star(t_f)\text{ if }t \geq t_f
    \end{cases}
\end{equation*}
where $0 \leq t_0 < t_f$, $T=(t-t_0)/(t_f-t_0)$. Note that $\dot{x}^\star(t)= \ddot{x}^\star(t)= \dddot{x}^\star(t)=0$ for $t \leq t_0$ and $t\geq t_f$. Our B\'ezier curve ensures continuity up to the third order derivative.
If $t > t_f > \delta$, $x(t_f) > \Delta$, the nominal value of the control variable is $R^\star (t) = C_0$. Choose for $R(t)$ the closest value $R_0^\star \in \mathfrak{U}$ to $C_0$. If $R_0^\star < C_0$ (resp. $R_0^\star > C_0$), $\dot{x} (t) > 0$ (resp. $\dot{x} (t) < 0)$ for $t$ large enough. Therefore $\lim_{t \to +\infty} x (t) = +\infty$ (resp. $x (t) < \Delta$ after some time). The necessity of closing the loop is obvious.




\subsection{Trajectory replanning}\label{Trajectoire}


To limit the bitrate variations, the trajectory is replanned according to the channel capacity. 
By imposing $\underline{x_t} \leq x(t) \leq \overline{x_t}$, $x(t)$ is alternatively increasing (${\tt dir}=1$) or decreasing (${\tt dir} = -1$):
IF {($x(t)>\overline{x_t}$ AND ${\tt dir}=1$)}, THEN ${\tt dir}=-1$;
IF {($x(t)<\underline{x_t}$ AND ${\tt dir}=-1$)}, THEN ${\tt dir}=1$.
When $x(t)$ is increasing (resp. decreasing), set ${\tt Coef} = \mathfrak{u}_{k}$ where $\mathfrak{u}_{k}$ is the element of $\mathfrak{U}$ immediately below (resp. above) $C_{\rm est}(t)$. 
Set
$$y^\ast_{\rm ad}(kT_e)=y^\ast_{\rm ad}((k-1)T_{\rm e})+(C_{\rm est}(kT_{\rm e})/{\tt Coef}-1)T_{\rm e}$$
where $T_{\rm e}$ is the sampling period. It yields a replanned trajectory that is expressed in discrete-time:
$y^\ast (kT_{\rm e}) + y^\ast_{\rm ad}(kT_{\rm e})$.
This replanning works well thanks to the bandwidth estimation in Section \ref{bandwidth}.



\subsection{Numerical values}
Set $t_0=0$s; $t_f=10$s; $x^\star(t_0)=0$; $x^\star(t_f)=4$; $\mathfrak{U} = \{0.332,0.6,1,2,3,5\}$; $C_0=0.7$;
$\alpha = - 10$ 
; $K_P = 0.25$; $\tau =1$s in the estimation of $C_{\rm est} (t)$; simulation duration: $10$min; $\delta = 5$s; $\Delta = 2$s and $u(t)$ is determined every $2$s;
sampling period: $T_{\rm e}=0.1$s.
\subsection{Scenarios}
\begin{figure*}[!ht]
\subfigure[\footnotesize Scenario 1]
{\epsfig{figure=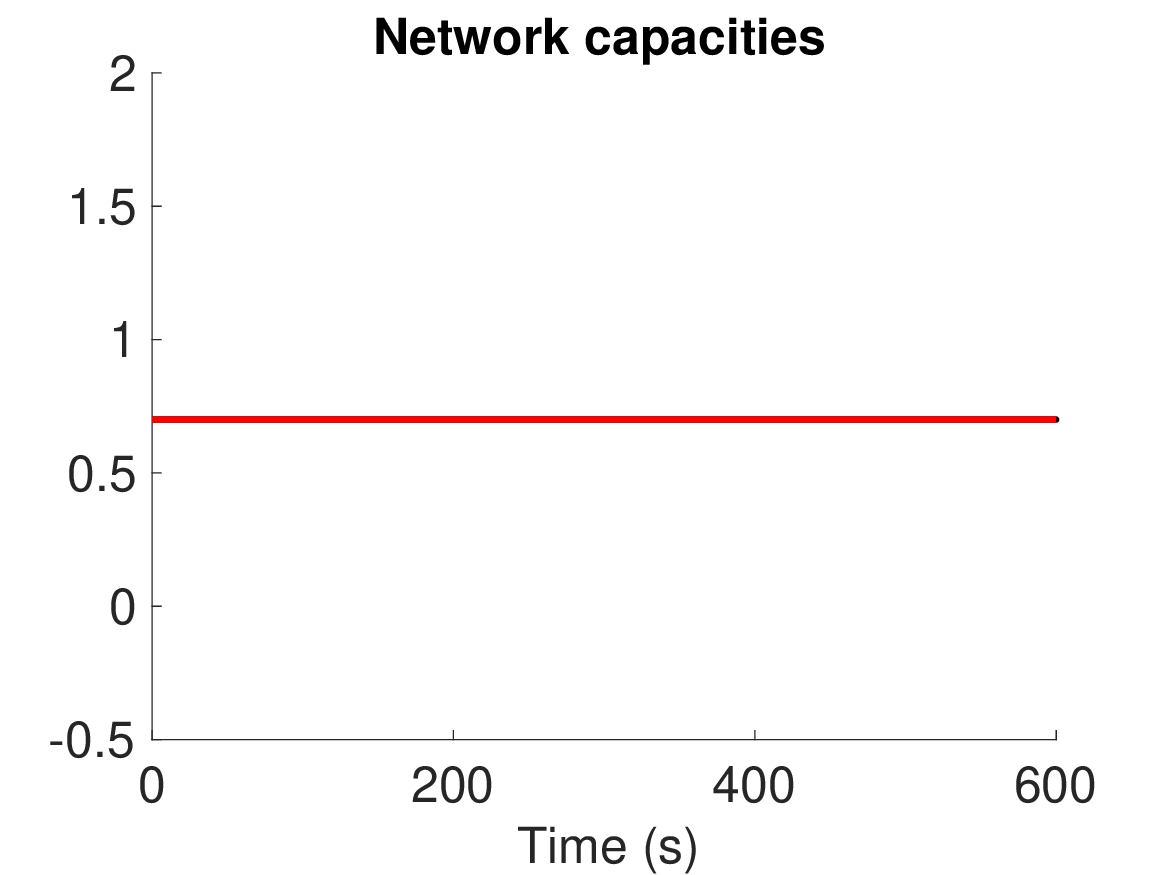,width=0.2891\textwidth}}
\subfigure[\footnotesize Scenario 2]
{\epsfig{figure=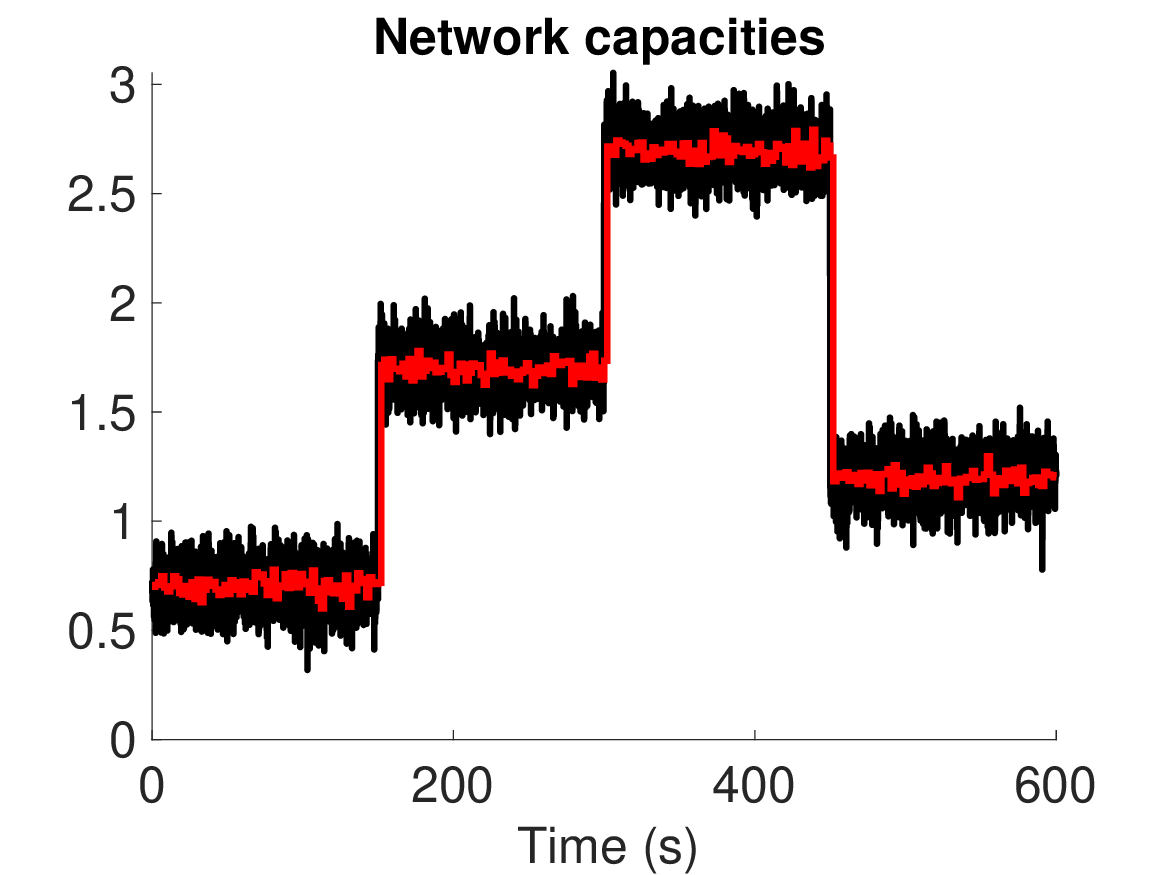,width=0.289\textwidth}}
\subfigure[\footnotesize Scenario 3]
{\epsfig{figure=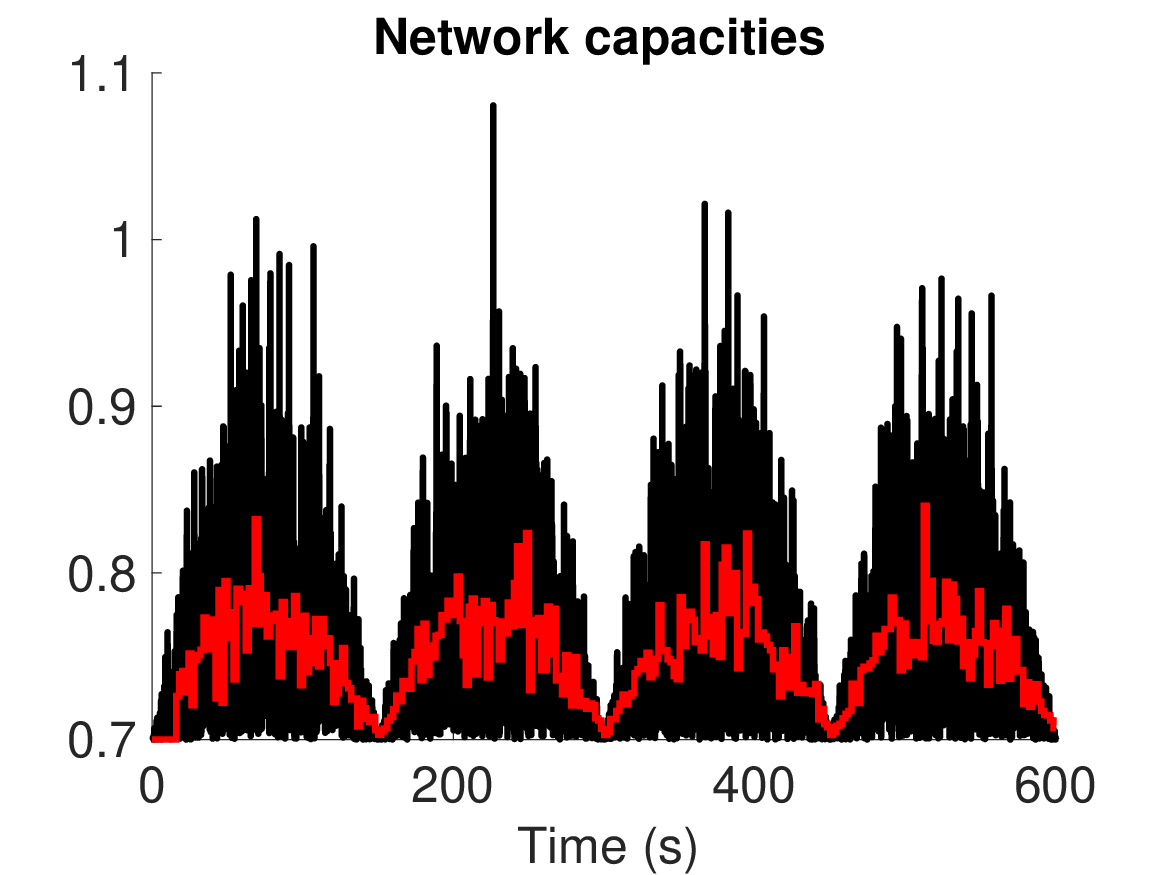,width=0.289\textwidth}}
\vspace{-0.2cm}\caption{Real channel capacity (black) and channel capacity estimation (red)}\label{C1a3}
\end{figure*}

\begin{figure*}[!ht]
\centering%
 \subfigure[\footnotesize Control (black) and open loop control (red) ]
{\epsfig{figure=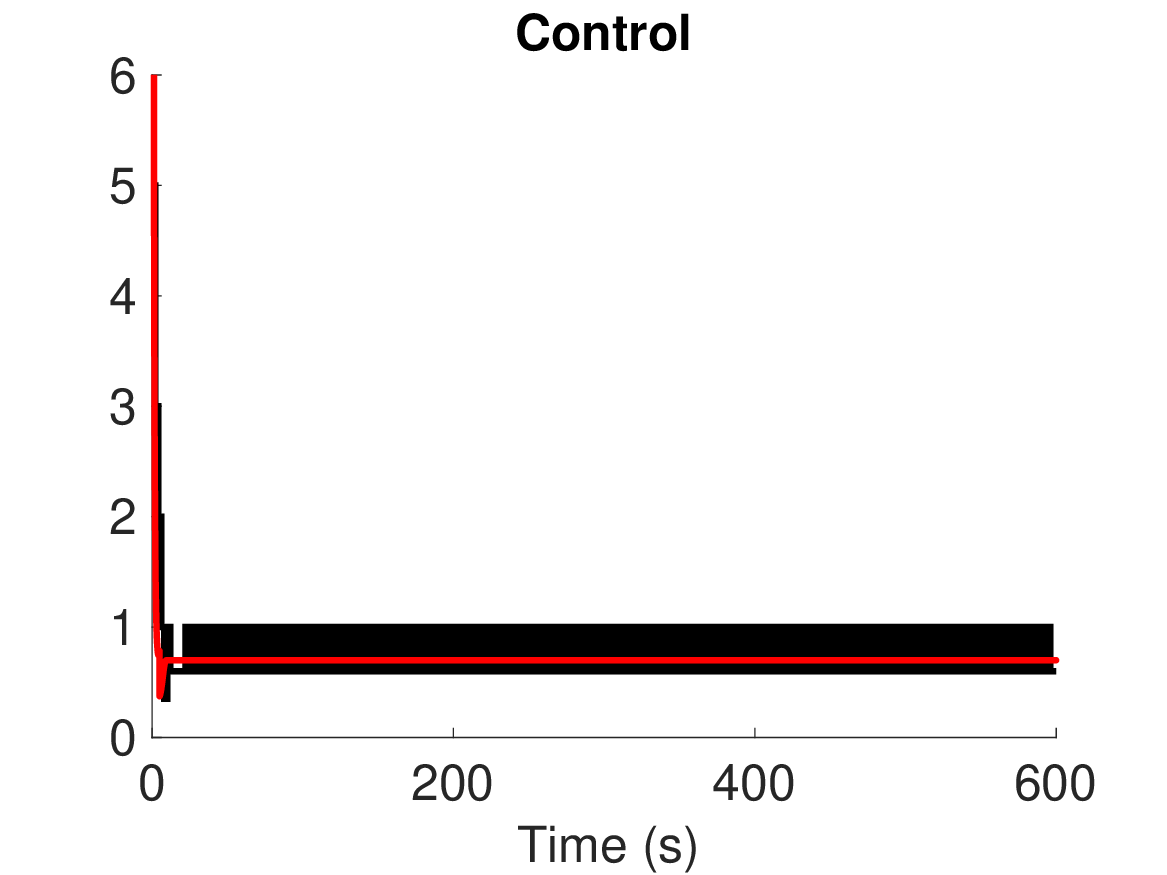,width=0.289\textwidth}}
\subfigure[\footnotesize Measured queue length (black), reference trajectory (red) and freeze threshold (green $- -$)]
{\epsfig{figure=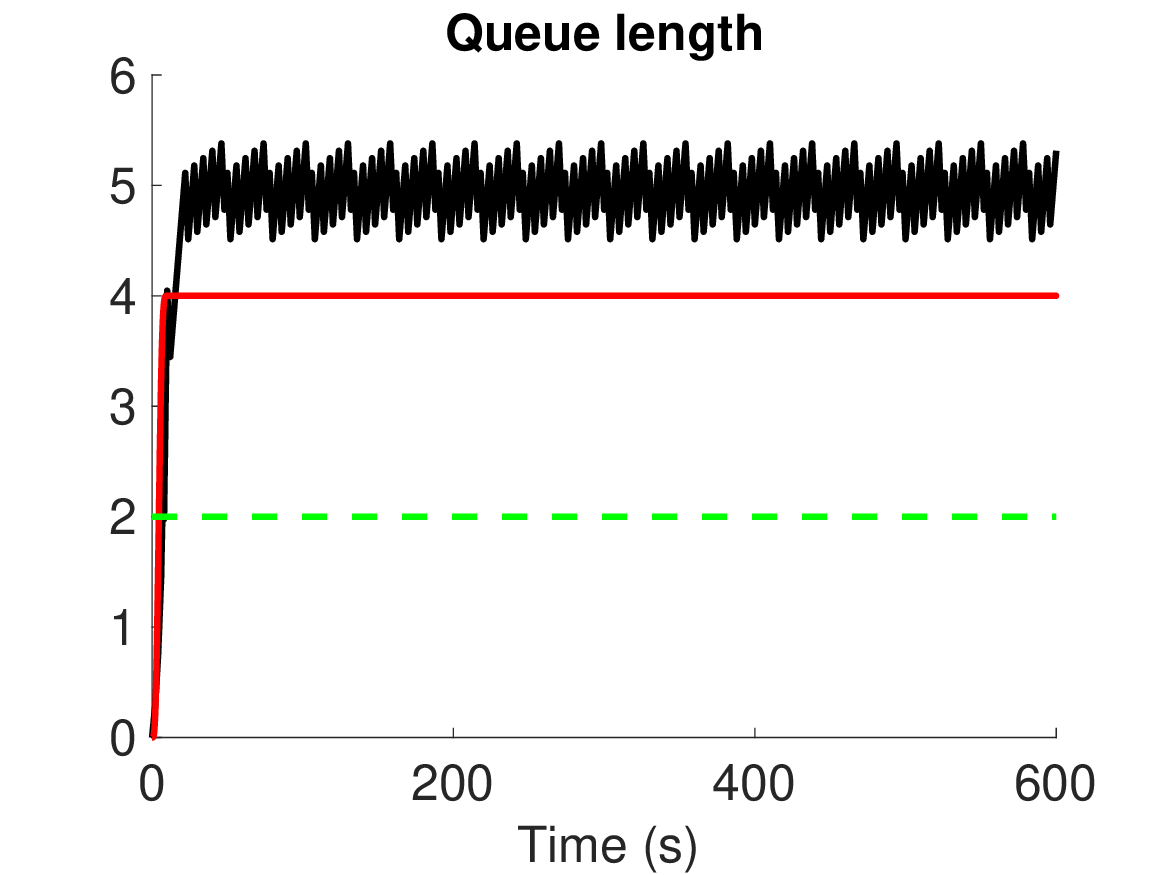,width=0.289\textwidth}}
\vspace{-0.2cm}\caption{Scenario 1 -- without trajectory replanning}\label{S11}
\end{figure*}

\begin{figure*}[!ht]
\centering%
\subfigure[\footnotesize Control (black) and open loop control (red) ]
{\epsfig{figure=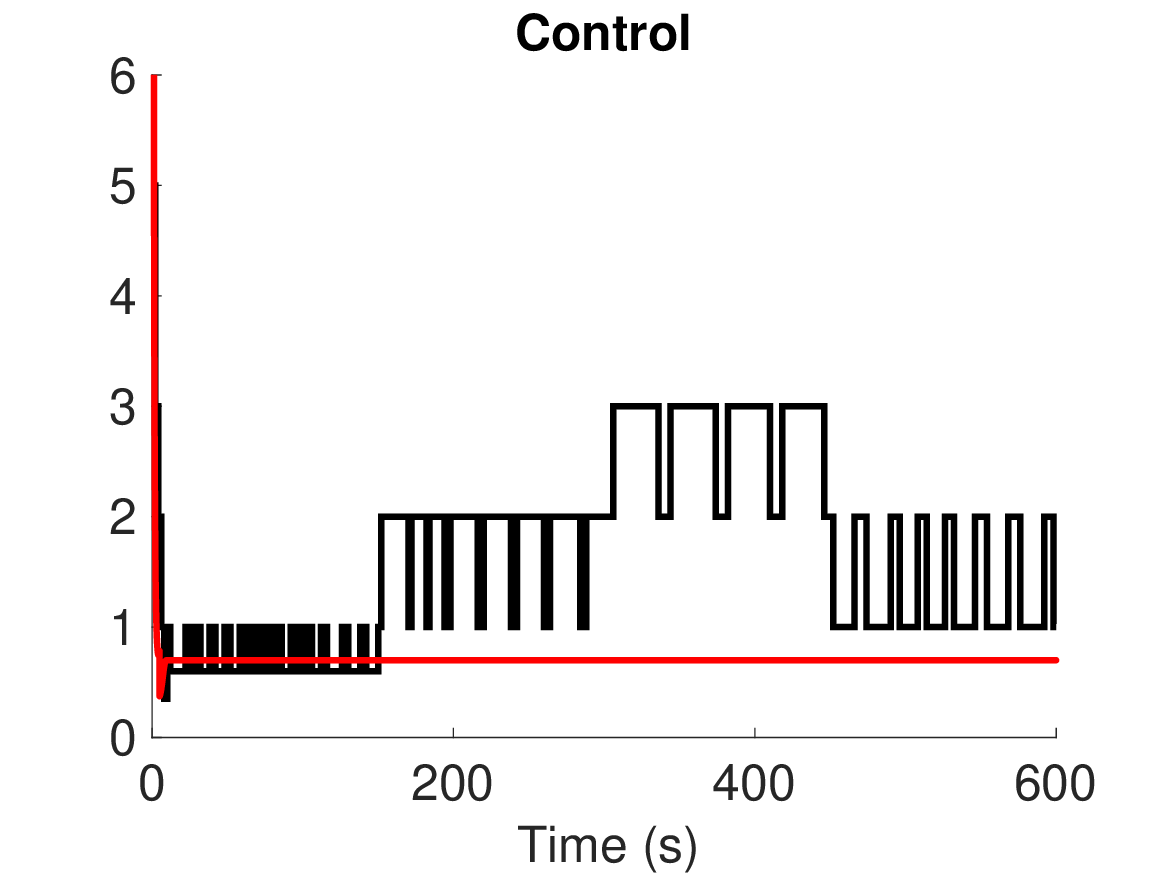,width=0.289\textwidth}}
\subfigure[\footnotesize Measured queue length (black), reference trajectory (red) and freeze threshold (green $- -$)]
{\epsfig{figure=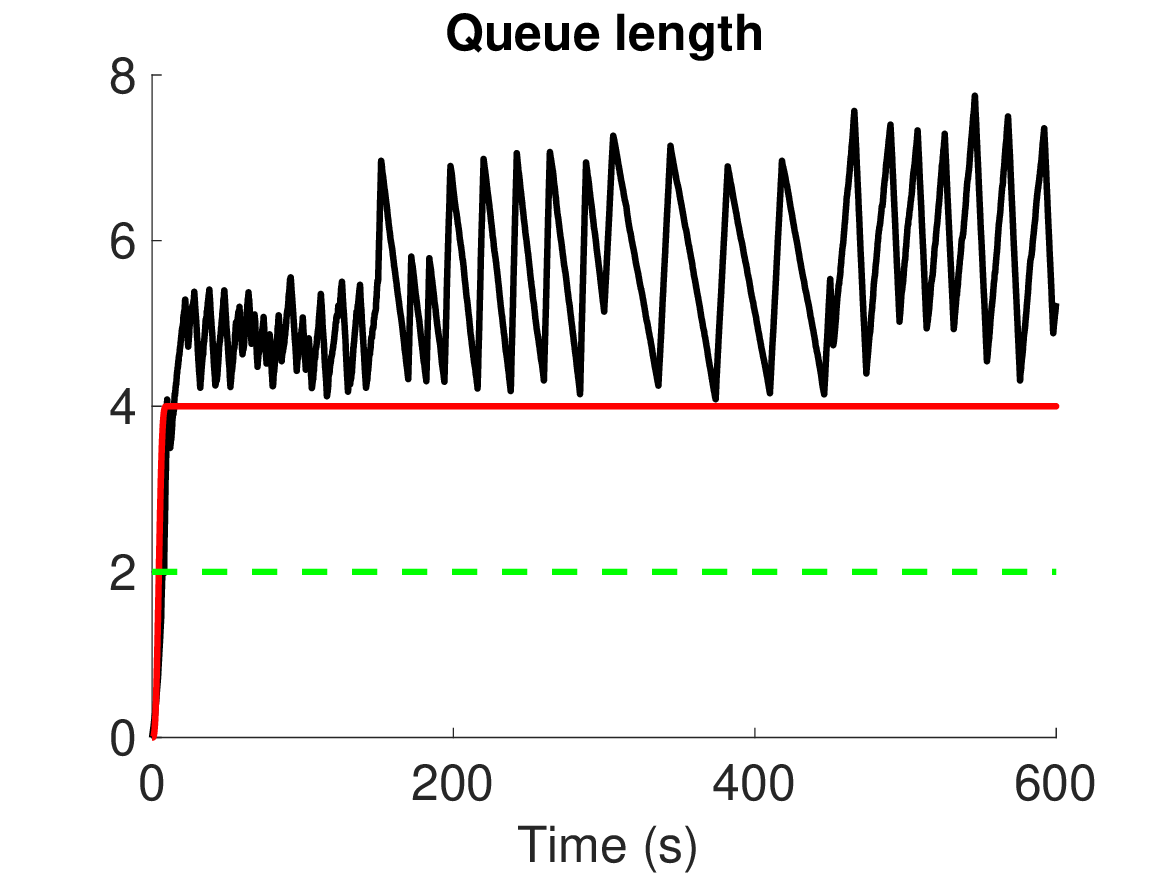,width=0.289\textwidth}}
\vspace{-0.2cm}\caption{Scenario 2 -- without trajectory replanning}\label{S13}
\end{figure*}

\begin{figure*}[!ht]
\centering%
\subfigure[\footnotesize Control (black) and open loop control (red) ]
{\epsfig{figure=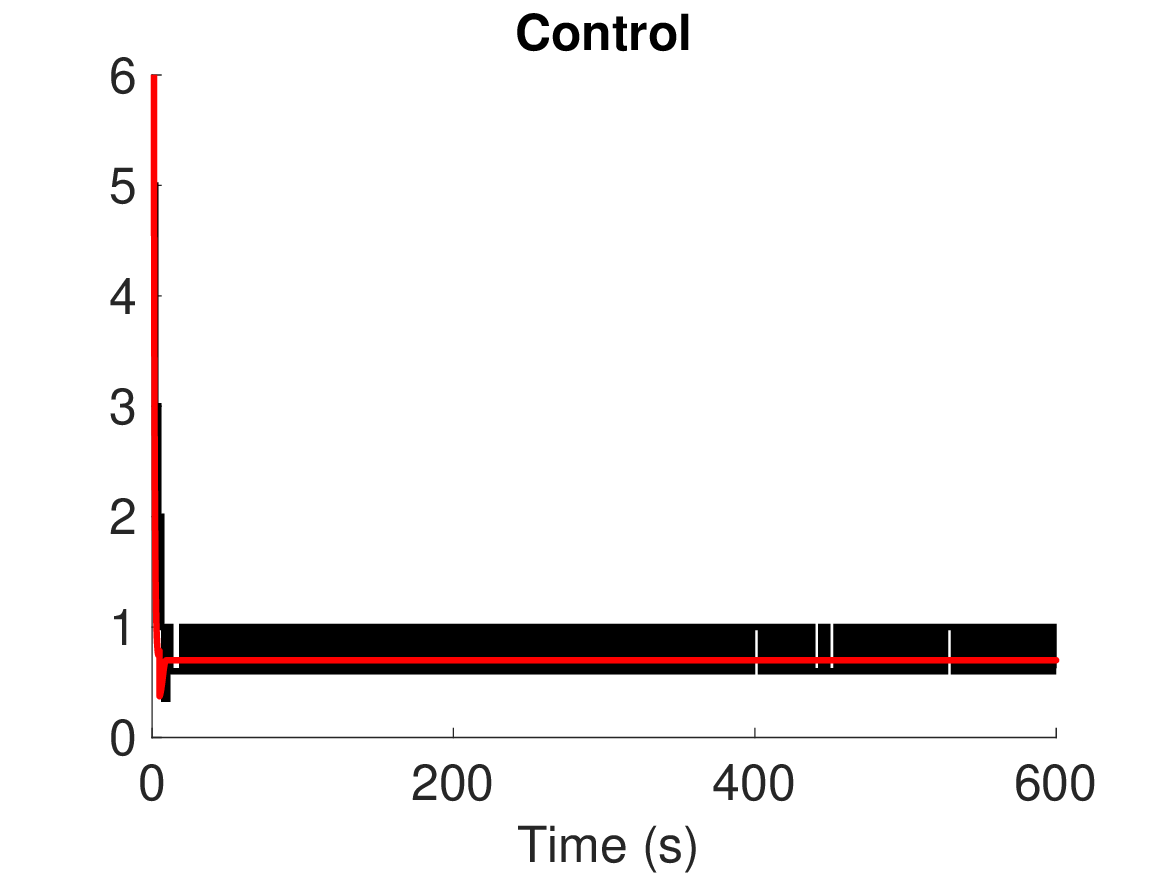,width=0.289\textwidth}}
\subfigure[\footnotesize Measured queue length (black), reference trajectory (red) and freeze threshold (green $- -$)]
{\epsfig{figure=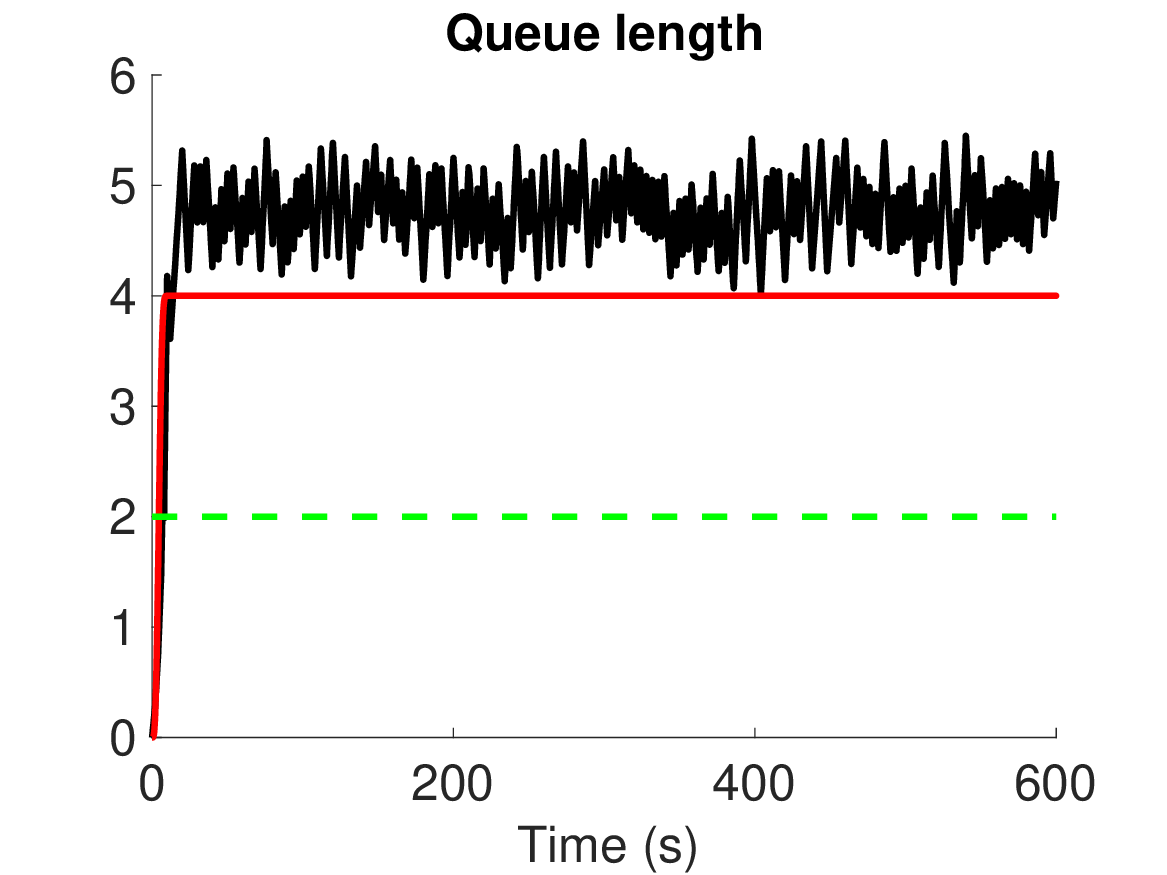,width=0.289\textwidth}}
\vspace{-0.2cm}\caption{Scenario 3 -- without trajectory replanning}\label{S14}
\end{figure*}

\begin{figure*}[!ht]
\centering%
\subfigure[\footnotesize Control (black) and open loop control (red) ]
{\epsfig{figure=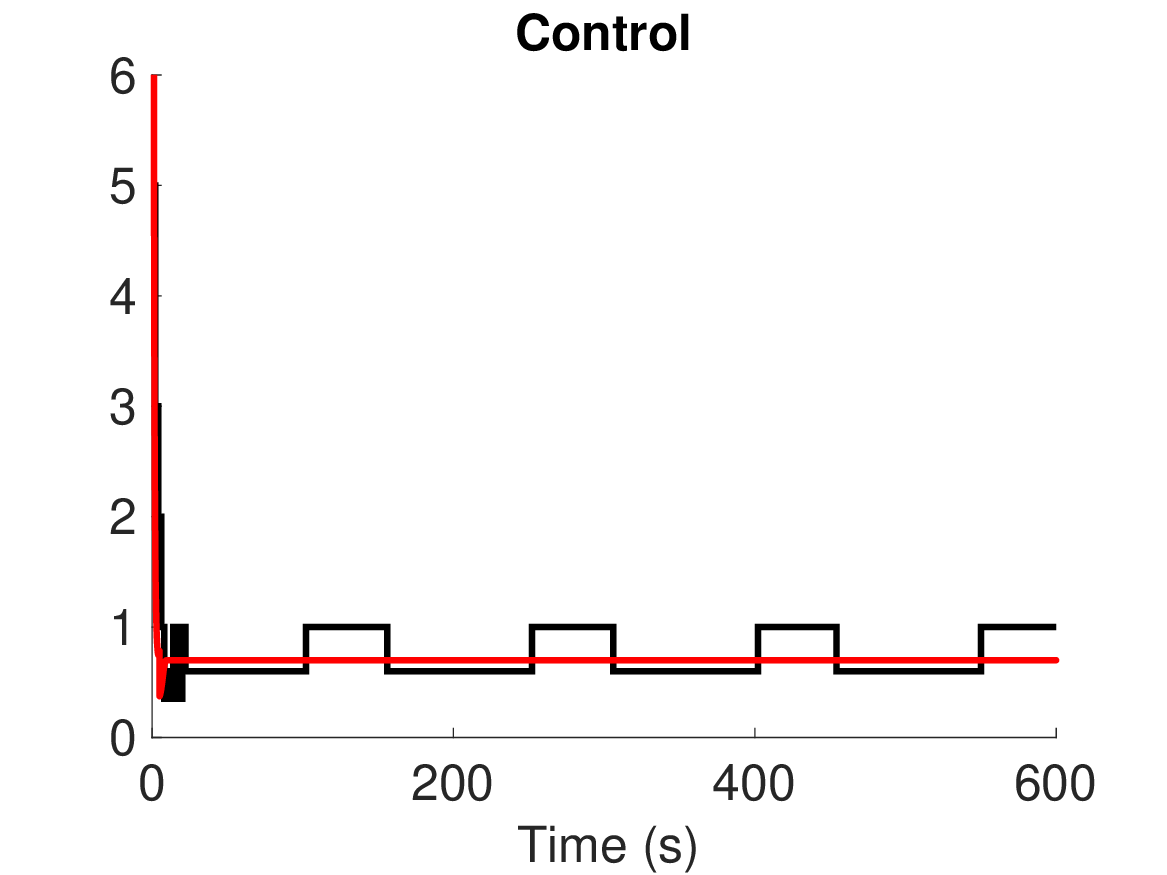,width=0.289\textwidth}}
\subfigure[\footnotesize Measured queue length (black), reference trajectory (red) and freeze threshold (green $- -$)]
{\epsfig{figure=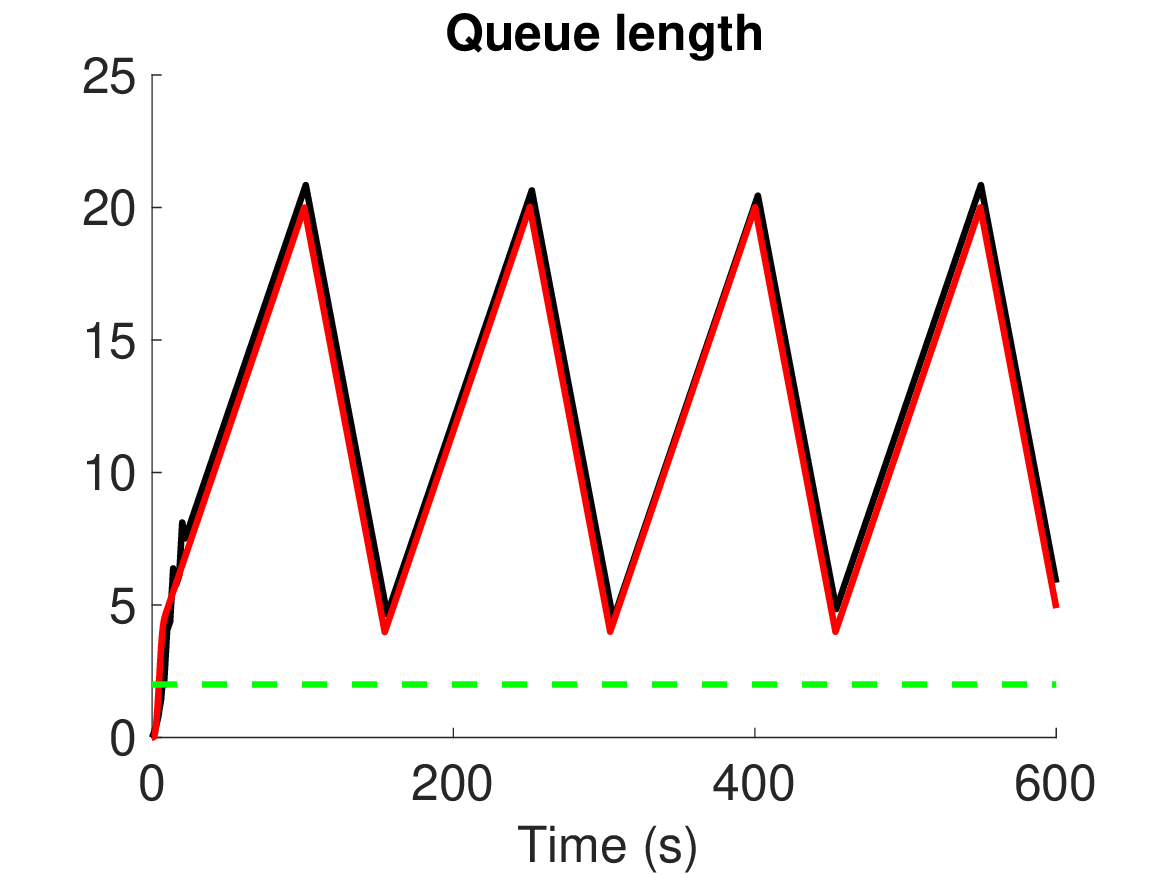,width=0.289\textwidth}}
\vspace{-0.2cm}\caption{Scenario 1 -- with trajectory replanning}\label{SE21}
\end{figure*}

\begin{figure*}[!ht]
\centering%
\subfigure[\footnotesize Control (black) and open loop control (red) ]
{\epsfig{figure=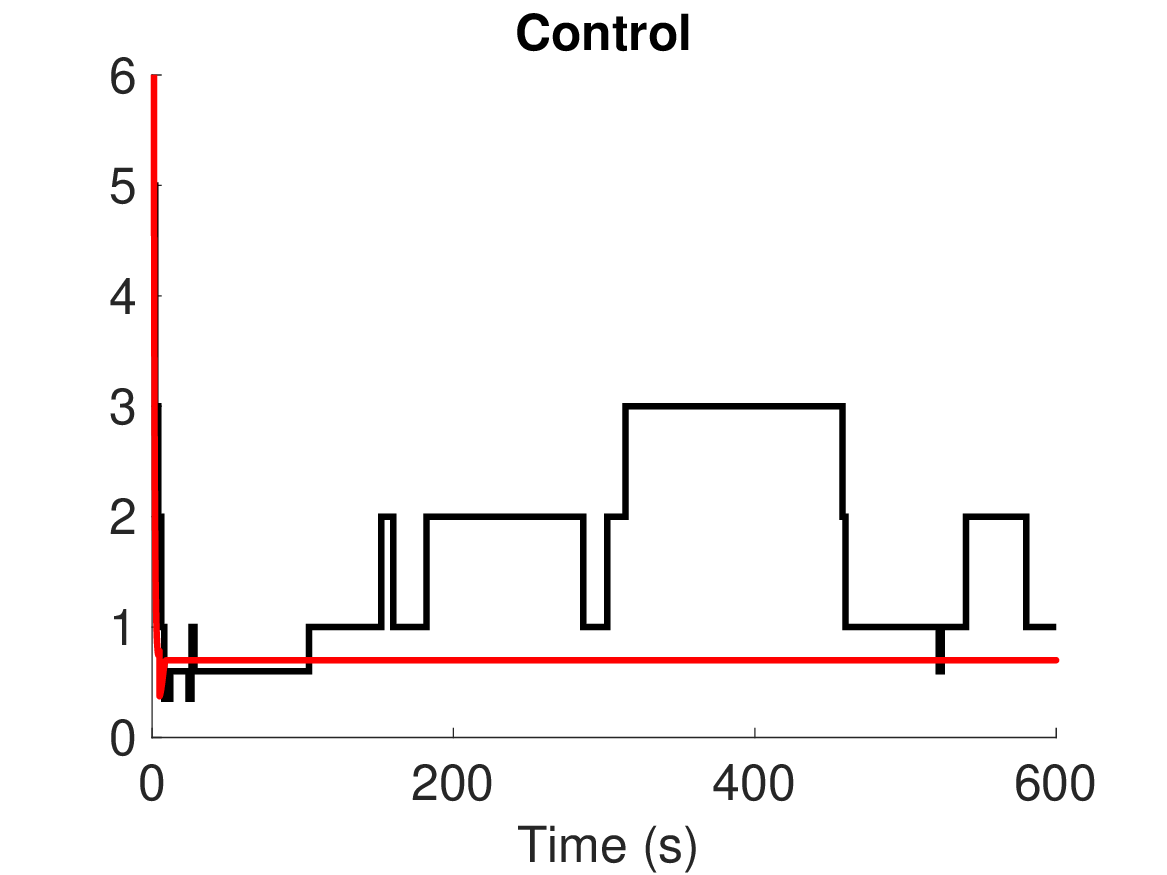,width=0.289\textwidth}}
\subfigure[\footnotesize Measured queue length (black), reference trajectory (red) and freeze threshold (green $- -$)]
{\epsfig{figure=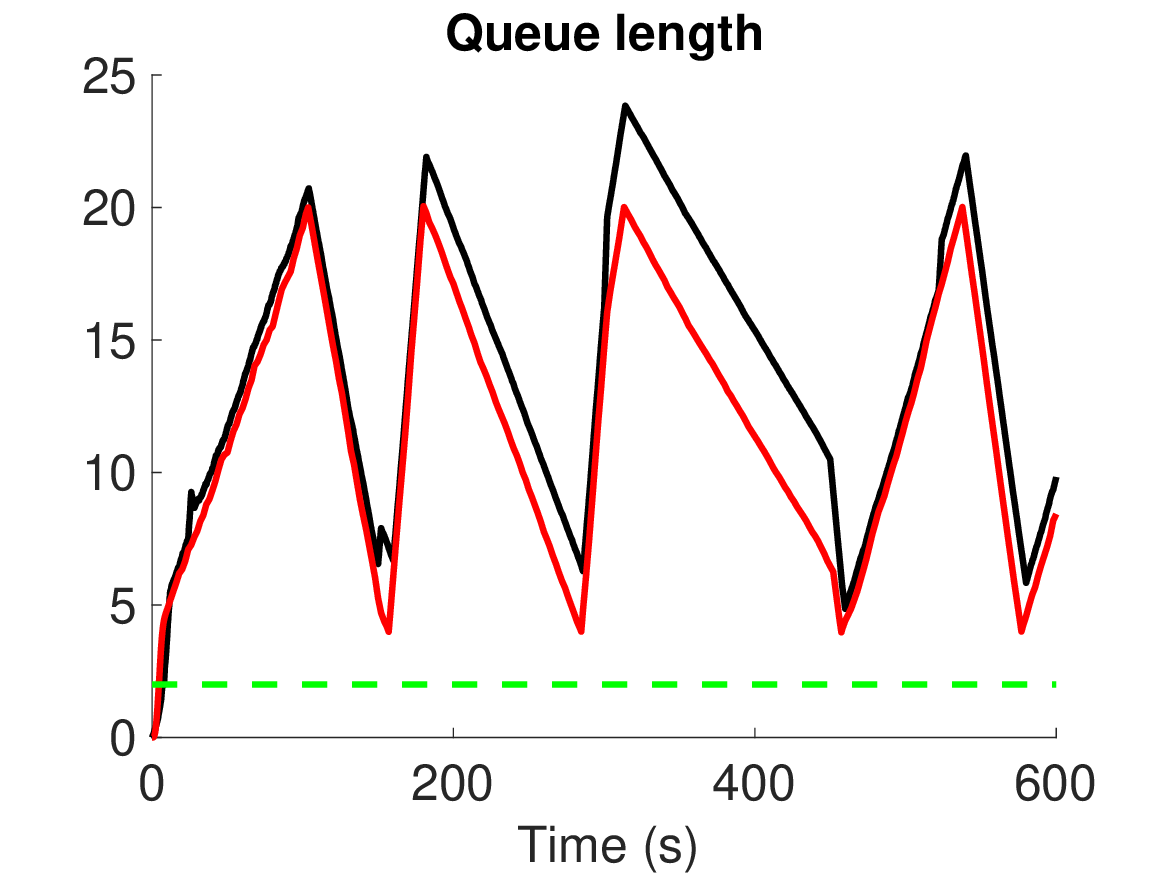,width=0.289\textwidth}}
\vspace{-0.2cm}\caption{Scenario 2 -- with trajectory replanning}\label{SE23}
\end{figure*}

\begin{figure*}[!ht]
\centering%
\subfigure[\footnotesize Control (black) and open loop control (red) ]
{\epsfig{figure=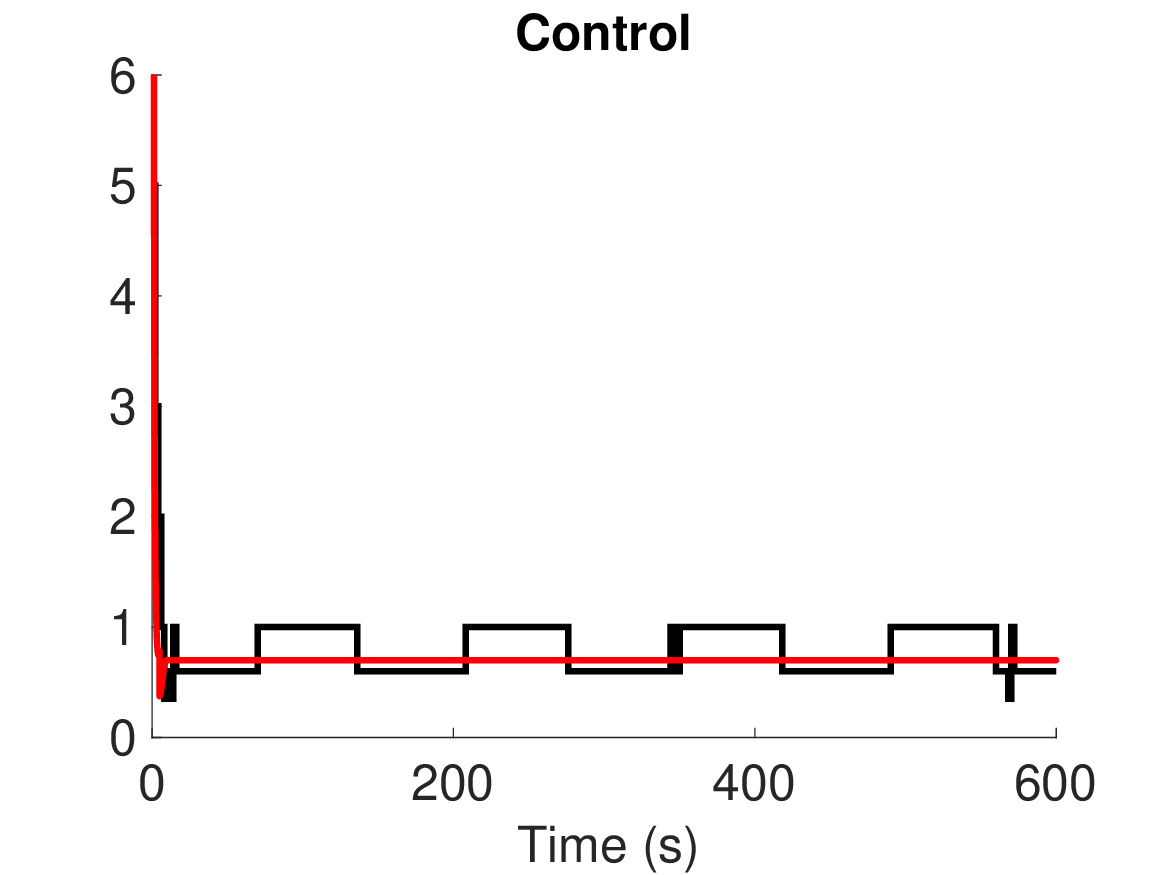,width=0.289\textwidth}}
\subfigure[\footnotesize Measured queue length (black), reference trajectory (red) and freeze threshold (green $- -$)]
{\epsfig{figure=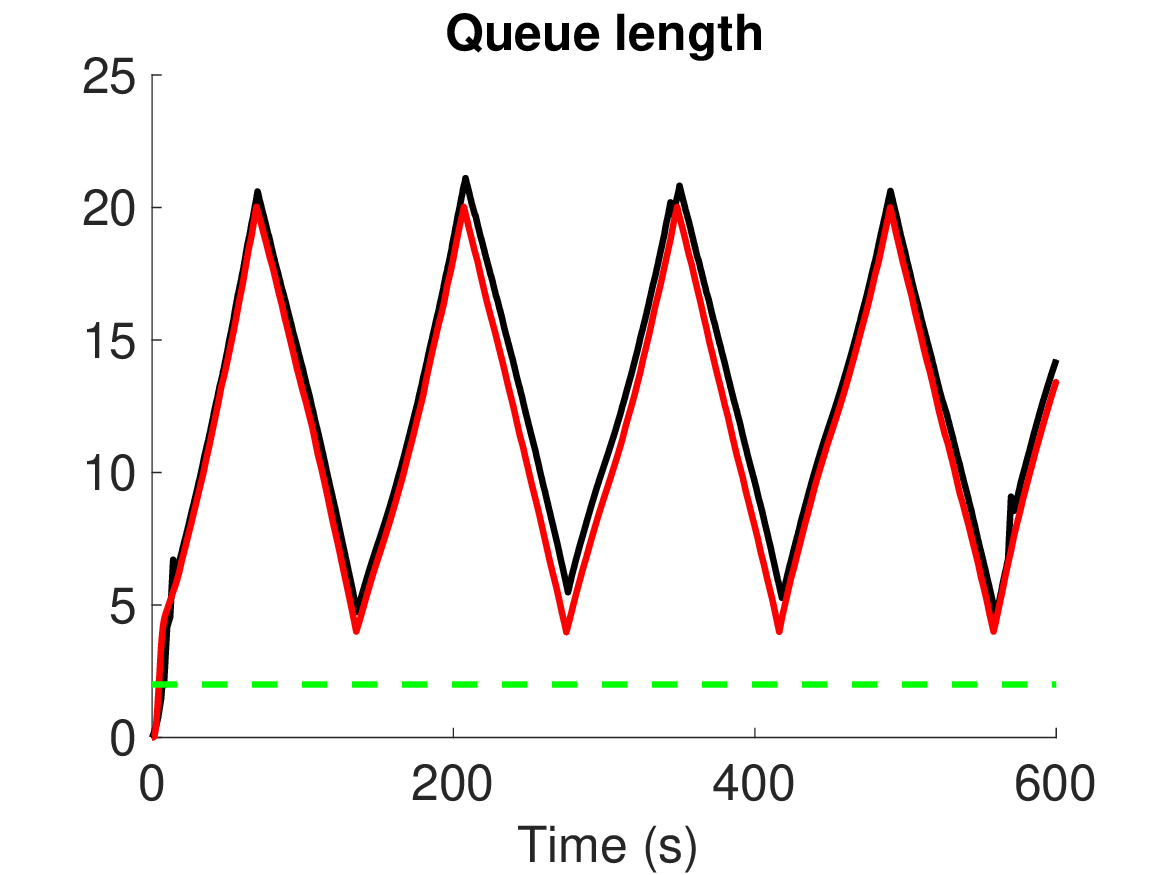,width=0.289\textwidth}}
\vspace{-0.2cm}\caption{Scenario 3 -- with trajectory replanning}\label{SE24}
\end{figure*}

\begin{itemize}
\item Scenario $1$: { $C(t)=C_0$} is perfectly known. See Figures \ref{C1a3}(a)-\ref{S11}-\ref{SE21}.
\item Scenario $2$: the measurement of $C(t)$, which is piecewise constant, is uncertain. See Figures \ref{C1a3}(b)-\ref{S13}-\ref{SE23}.
\item Scenario $3$:  strong variations and uncertainties for $C(t)$. See Figures \ref{C1a3}(c)-\ref{S14}-\ref{SE24}.
\end{itemize}

\subsection{Evaluation metrics}\label{metric}
Introduce (compare with \cite{qin0,qin,qin.phd}, and references therein) three evaluation metrics for a video of $M$ chunks, where $R(t_k)$ is the bitrate for the k$th$ chunk, which starts being downloaded at time $t_k$:
\begin{enumerate}
    \item \emph{Average video quality}: $\frac{1}{M} \sum_{k = 1}^{M} R(t_k)$.
    \item \emph{Average quality variation}: $$\frac{1}{M - 1} \sum_{k = 1}^{M-1} |\mathbf{sign}(R(t_{k+1})- R(t_k))|$$ where $\mathbf{sign}$ is the sign function, \textit{i.e.}, $\mathbf{sign}(\sigma) = 1$ if $\sigma > 0$, 
    $\mathbf{sign}(\sigma) = 0$ if $\sigma = 0$, 
   $\mathbf{sign}(\sigma) = - 1$ if $\sigma < 0$,
    \item \emph{Rebuffering time}: $\sum_{k = 1}^{M} \mathbf{H}(\Delta - x(t_k))$, where $\mathbf{H}$ is the Heaviside function, \textit{i.e.}, $\mathbf{H} (\sigma) = 1$ if $\sigma \geq 0$, $\mathbf{H} (\sigma) = 0$ if $\sigma < 0$.
\end{enumerate}
As well known (see, e.g., \cite{Knuth}), the computational power required to execute an algorithm depends on the number of elementary operations performed. Table I thus shows very low computing power.
Tables II and III use 100 simulations to demonstrate the undeniable benefits of replanning. The \emph{average quality variation} is reduced by up to 4 times, depending on the scenario. Let us stress that the \emph{average video quality} is not altered.

\begin{table*}[!ht]
\begin{center}
\caption{Complexity of calculations}\label{tb5}
\begin{tabular}{|c|cccc|}
\hline
Functions & $\times$ and $/$ & $+$ and $-$ & assignments & test functions  \\\hline\hline
Capacity estimation &25&10&2&1\\
Trajectory replanning&2&10&10&10\\
MFC &50&50&30&5\\
Others&0&5&20&10\\\hline\hline
In total &77&75&62&26\\\hline
\end{tabular}
\end{center}
\end{table*}

The discontinuous nature of the control variable explains the imperfection of reference trajectory tracking. Increasing $K_p$ would improve it but could lead to more changes in quality, which might be unpleasant for the viewer. 

\section{Conclusion}\label{conclu}
Our flatness-based approach could help overcome some difficulties with classic sampling techniques \cite{lim}. We propose to estimate bandwidth only via usual measurements, contrary to what is typically done.
Careful comparisons with other approaches are, of course, necessary, as well as validation through real-world network experiments (see, \textit{e.g.}, \cite{acm,algo}) in order to fully validate our viewpoint. Let us finally emphasize again that the computational burden of our algebraic calculations is low.

\begin{table*}[!ht]
\begin{center}
\caption{QoE without trajectory replanning -- $x^\star(t_f)=4$ -- $\left(x^\star(t_f)=12\right)$}\label{tb1}
\begin{tabular}{|c|cccc|}
\hline
Network capacities & Average video quality & Average quality variation & Rebuffering time & Figures  \\\hline\hline
Scenario 1 &0.7408 (0.7394)&206 (206)&0 (0)& \ref{S11}\\
Scenario 2 &1.6971 (1.6972)&73.70 (73.15)&0 (0)& \ref{S13}\\
Scenario 3 &0.8 (0.7985)&212 (217.08)&0 (0)& \ref{S14}\\\hline
\end{tabular}
\end{center}
\end{table*}


\begin{table*}[!ht]
\begin{center}
\caption{QoE with trajectory replanning }\label{tb4}
\begin{tabular}{|c|cccc|}
\hline
channel capacities & Average video quality & Average quality variation & Rebuffering time & Figures   \\\hline\hline
Scenario 1 &0.7461&10.54&0& \ref{SE21}\\
Scenario 2 &1.7172&23.64&0& \ref{SE23}\\
Scenario 3 &0.8027&16.56&0& \ref{SE24}\\\hline
\end{tabular}
\end{center}
\end{table*}



\end{document}